\documentclass[proof]{WileyASNA-v1}

\articletype{Original Article}%

\received{30 August 2024}
\revised{XX YY ZZZZ}
\accepted{XX YY ZZZZ}

\def\lapp{\ifmmode\stackrel{<}{_{\sim}}\else$\stackrel{<}{_{\sim}}$\fi}
\def\gapp{\ifmmode\stackrel{>}{_{\sim}}\else$\stackrel{>}{_{\sim}}$\fi}

\begin{document}

\title{Revisiting the Intrabinary Shock Model for Millisecond Pulsar Binaries: Radiative Losses and Long-Term Variability}

\author[1]{Jaegeun Park}

\author[1]{Chanho Kim}

\author[1]{Hongjun An}

\author[2,3,4]{Zorawar Wadiasingh}

\authormark{Jaegeun Park \textsc{et al}}

\address[1]{\orgdiv{Department of Astronomy and Space Science}, \orgname{Chungbuk National University}, \orgaddress{\state{Cheongju 28644}, \country{Republic of Korea}}}
\address[2]{\orgdiv{Department of Astronomy}, \orgname{University of Maryland College Park}, \orgaddress{\state{MD 20742}, \country{USA}}}
\address[3]{\orgdiv{Astrophysics Science Division}, \orgname{NASA GSFC}, \orgaddress{\state{MD 20771}, \country{USA}}}
\address[4]{\orgdiv{Center for Research and Exploration in Space Science and Technology}, \orgname{NASA GSFC}, \orgaddress{\state{MD 20771}, \country{USA}}}

\corres{*Hongjun An, Department of Astronomy and Space Science, Chungbuk National University, Cheongju 28644, Republic of Korea. \email{hjan@cbnu.ac.kr}}

\abstract{Spectrally hard X-ray emission with double-peak light curves (LCs) and orbitally modulated gamma rays
have been observed in some millisecond pulsar binaries,
phenomena attributed to intrabinary shocks (IBSs). While the
existing IBS model by \citet{Sim+2024} successfully explains these high-energy features
observed in three pulsar binaries,
it neglects particle energy loss within the shock region.
We refine this IBS model to incorporate radiative losses of X-ray emitting electrons and positrons, and
verify that the losses have insignificant
impact on the observed LCs and spectra of the three binaries. 
Applying our refined IBS model to the X-ray bright pulsar binary PSR~J1723$-$2837, we predict that it can be detected by the Cherenkov Telescope Array.
Additionally, we propose that the long-term X-ray variability observed in XSS~J12270$-$4859 and PSR~J1723$-$2837 is due to changes in the shape of their IBSs.
Our modeling of the X-ray variability suggests that these IBS shape changes may alter the extinction of the companion's optical emission, potentially explaining the simultaneous optical and X-ray variability observed in XSS~J12270$-$4859.
We present the model results and discuss their implications.
}




\maketitle

\section{Introduction}\label{sec1}
Millisecond pulsar binaries, hosting a millisecond pulsar and a low-mass companion, form after a long period
of accretion from the companion ceases.
This accretion phase spins up the neutron star to millisecond
periods \citep[][]{acrs82} and can significantly increase its mass \citep[e.g.,][]{vanKerkwijk2011}.
Therefore, pulsar binaries serve as invaluable probes for studying the evolution of low-mass X-ray binaries
and constraining the equation of state of dense matter \citep[e.g.,][]{Linares2020}.
To fully exploit their potential, a comprehensive understanding of the pulsar-companion interaction
and accurate determination of system parameters, such as orbital inclination,
through broadband emission modeling are essential.

\begin{figure*}[ht]
\centerline{
\begin{tabular}{ccc}
\includegraphics[width=2.06 in]{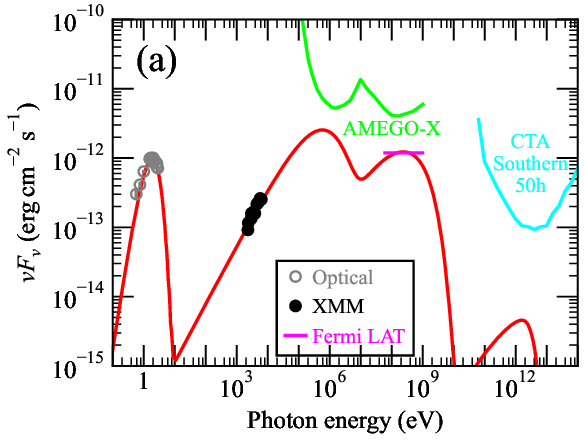} &
\includegraphics[width=1.98 in]{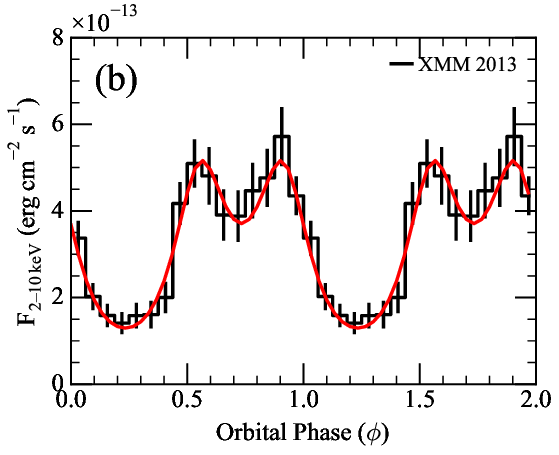} &
\includegraphics[width=2.07 in]{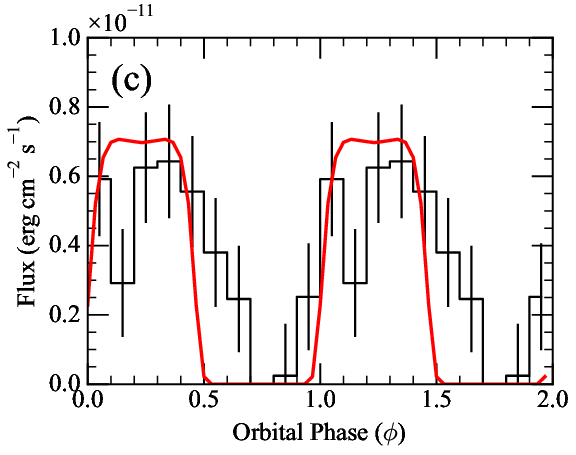} 
\vspace{-6mm}
\end{tabular}}
\caption{
\scriptsize{Spectral energy distribution (SED) and light curves (LCs) of XSS~J12270$-$4859 \citep[][]{Sim+2024}.
(a): SED displaying blackbody emission from the companion (gray), and non-thermal X-ray data (black).
Estimated flux of the orbitally modulated $\lapp$GeV signal
is indicated by the pink line. Green and cyan curves show the sensitivity of the AMEGO-X and CTA observatories.
(b): X-ray LC in the 2--10\,keV band measured by XMM-Newton.
(c) 100\,MeV--1\,GeV LC measured by Fermi LAT \citep[][]{An2022}. The red curves present our model (Table~\ref{tab1})}.
\vspace{-7mm}
\label{fig1}}
\end{figure*}

Observationally, compact millisecond pulsar binaries are categorized into black widows (BWs)
and redbacks (RBs) depending on the companion's mass
($M_c$); BWs have $M_c<0.1M_\odot$, and RBs have $M_c>0.1M_\odot$
\citep[see][]{Strader2019,Swihart2022}.
BWs and RBs share similar emission properties, most notably
the prominent orbital modulation across multiple wavelengths.
Some systems undergo radio eclipses caused by the companion's outflow at specific
orbital phases. In the optical band, the characteristic day-night cycles arise from the pulsar's
heating of the companion and ellipsoidal modulation. 
These systems display hard power-law (PL) X-ray emission
(photon index $\Gamma\approx 1.2$) with orbital modulation,
typically characterized by a single or double-peak light curve (LC).
The X-ray emission peaks around orbital phase
$\phi\approx 0.75$ or $\phi\approx 0.25$ depending on the IBS orientation, where
$\phi\approx 0.75$ denotes the phase when the companion is behind the pulsar as seen by the observer.
These phenomena allow us to study the pulsar wind and
its interaction with the companion's wind or magnetic field \citep[e.g.,][]{Harding1990,Wadiasingh2018}.

Previously, gamma-ray emission from these systems
was thought to arise from the pulsar magnetosphere (i.e., pulsed).
However, recent Fermi-LAT detections of orbitally modulated $\lapp$GeV signals in a few systems
challenge this notion \citep[e.g.,][]{ark18,ntsl+18}. These signals reach $\approx$30\%
of the pulsed magnetospheric flux, with a LC maximum at $\phi\approx 0.25$. Moreover, there is evidence suggesting that these orbitally modulated signals may be pulsed and have spectra distinct from the pulsar's emission \citep[][]{ark20,Clark+2021}.
A comprehensive study of this high-energy emission, employing a suitable emission model,
could provide crucial insights into the energetic processes within these systems.
Such an investigation might shed light on the origin of the recently detected positron excess
\citep[][]{Linares2021} and elucidate the physical mechanisms operating in more energetic
TeV gamma-ray binaries \citep[e.g.,][]{Kim+2022}.

This paper investigates the high-energy emission properties of RBs.
Section~\ref{sec2} presents a model for high-energy emission in these systems.
In Section~\ref{sec3}, we apply our IBS model to
the LCs and spectral energy distributions (SEDs) of a sample of RBs (Section~\ref{sec3_1}).
Additionally, we model the long-term X-ray variability observed
in PSR~J1723$-$2837 and XSS~J12270$-$4859 (J1723 and J1227 hereafter; Section~\ref{sec3_2}).
Finally, Section~\ref{sec4} discusses our results.

\section{A Model for the X-ray and Gamma-ray Emission}\label{sec2}
Previous IBS models \citep[e.g.,][]{kra19} primarily focus on X-ray emission.
These models postulate that relativistic electrons and positrons
(hereafter referred to as ``electrons'') are accelerated by the IBS.
These particles flow along the shock, emitting radiation that is Doppler-boosted in the direction
of their motion, creating a bright ring-like emission region \citep[e.g.,][]{Romani2016}.
When the observer's line of sight (LoS) intersects
this ring, an X-ray intensity peak is observed. The X-ray LC peak at $\phi\approx 0.75$
commonly observed for RBs (Fig.~\ref{fig1} (b)) suggests that the IBS is bent towards the pulsar
in RBs, leading to the strongest Doppler beaming at this orbital phase \citep[e.g.,][]{Wadiasingh2017}.

Orbitally modulated GeV emission from RBs remains poorly understood.
Proposed origins include the pre-shock pulsar wind and the IBS.
Synchrotron radiation from the IBS struggles to reproduce the observed
GeV LC phasing, while inverse-Compton (IC) emission from the IBS
predicts a peak in the SED at TeV energies,  inconsistent with observations.
Alternatively, IC emission from the pre-shock pulsar
wind could explain the GeV LC shape, but this component cannot
explain the observed flux levels \citep[][]{ark20,Clark+2021}.
Furthermore, these scenarios cannot account for the possible ``pulsations'' in the orbitally-modulated emission since their emission timescales are longer than the pulse periods of a few milliseconds.

To address the challenges in explaining orbitally modulated GeV emission,
\citet{Merwe+2020} proposed a scenario where high-energy electrons from the pulsar
wind have a large gyro radius (greater than the IBS thickness), traverse the IBS, and interact with the companion's strong magnetic field ($B$)
of $\sim$\,kG. \citet{Sim+2024} (SAW24 hereafter) adopted this `shock-penetration' scenario.
In this framework, rapid energy losses of high-energy electrons ($\propto B^2 E_e^2$)
due to the companion's strong $B$ is the key factor that generates
$\lapp$GeV emission and its orbital modulation.
Consequently, SAW24 carefully considered cooling of the shock-penetrating electrons.
While successfully reproducing observational data for three RBs,
the SAW24 model neglected radiative cooling of electrons within the IBS itself,
assuming that it is insignificant for these systems.
However, such cooling could potentially modify the energy distribution of
the highest-energy electrons within the IBS, thereby affecting the resulting X-ray spectrum and
associated IC emission in the TeV band.

To incorporate radiative cooling within the IBS, we refined the multi-zone IBS model of SAW24.
Unlike SAW24, which assumed a stationary electron distribution throughout the IBS,
we dynamically updated the distribution in each zone to account for radiative cooling.
Cooled particles were then transferred to the subsequent emission zone,
and this process was iterated along the electron flow through the IBS.

\section{Application of the model}\label{sec3}
We apply our updated model to the three RBs used by SAW24, as well as to the X-ray bright
RB J1723 (Section~\ref{sec3_1}).
Additionally, we interpret the long-term X-ray variability observed in 
J1227 and J1723 within the framework of our IBS model (Section~\ref{sec3_2}).
We then propose a scenario to explain optical variability in J1227 that was contemporaneous
with the long-term X-ray variability (Section~\ref{sec3_3}).

\subsection{Broadband SEDs and Multi-band LCs}\label{sec3_1}
Figure~\ref{fig1} presents model results for J1227 as an example. 
Our results for the three RBs modeled by SAW24 are consistent with their findings,
suggesting that radiative cooling within the IBS has a negligible impact on
the observed LCs and SEDs.
This is likely due to the limited energy range of current high-energy observatories,
which do not cover energies where the most energetic electrons predominantly emit.
A high-quality spectrum extending to higher energies \citep[e.g., by AMEGO-X;][]{AMEGOX2022}
might reveal the cooling effect.

\begin{figure}[ht]
\centerline{\includegraphics[width=2.75 in]{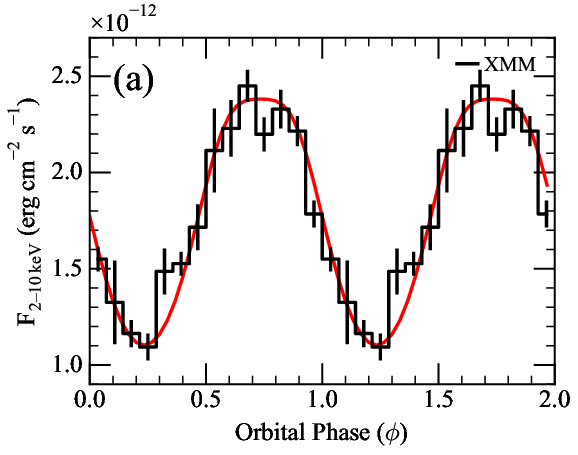}}
\vspace{-1 mm}
\hspace{-3.5mm}
\centerline{\includegraphics[width=2.8 in]{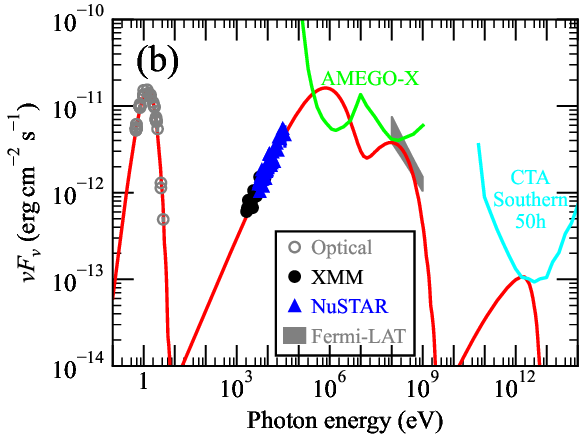}}
\vspace{-4mm}
\caption{
\scriptsize{X-ray LC and broadband SED of J1723.
(a) X-ray LC obtained with XMM-Newton in 2011 \citep[][]{Bogdanov+2014}.
(b) Companion emission is represented by gray points.
X-ray spectra were measured by XMM-Newton (black circles) and NuSTAR (blue triangles) \citep[][]{Kong+2017}, and Fermi-LAT data are shown as a gray band \citep[][]{Hui+2014}.
Red curves in both panels represent our model computations (Table~\ref{tab1}), while the sensitivity curves
for AMEGO-X and CTA are as presented in Figure~\ref{fig1}.}
\vspace{-6mm}
\label{fig2}}
\end{figure}

A particularly interesting case is the bright RB J1723.
Its X-ray spectrum, measured by NuSTAR up to 79\,keV, exhibits a single PL
with $\Gamma=1.3$. Its LC displays a single peak at
$\phi=0.75$ \citep[][]{Kong+2017}.
While GeV emission from this source has been reported
\citep[][]{Hui+2014}, its origin remains uncertain due to absence of
detected pulsations or orbital modulation.\footnote{This source
is not listed in the 4FGL catalog, making the claimed LAT detection less secure.}
\citet{Merwe+2020} proposed two possible scenarios for this source.
In the first, synchrotron radiation from the IBS produces only X-rays,
assuming the reported GeV detection is spurious.
Alternatively, the IBS could generate both X-ray and GeV photons,
requiring strong Doppler beaming for synchrotron emission in the GeV band.
A similar scenario has been proposed for the BW PSR~J1311$-$3430 \citep[][]{arjk+17}.

Given the similarities in the X-ray SED and LC phasing between J1723 and other RBs,
we employed the scenario where X-rays originate from the IBS and gamma rays from the companion zone.
Figure~\ref{fig2} presents the model results.
To explore a limiting case, we assumed no magnetospheric contribution to
the GeV flux. This assumption directly influences the parameters for shock-penetrating electrons
and the companion's $B$. In reality, the magnetospheric component may contribute up to
$\sim$70\% of the GeV flux, as observed in other RBs, requiring a correspondingly
reduced contribution from shock-penetrating electrons.
Our model can accommodate lower gamma-ray flux, including nondetection,
by adjusting the number of penetrating particles or the companion's $B$,
without altering the IBS or pre-shock parameters (Table~\ref{tab1}).

\newcommand{\marka}{\tablenotemark{a}}
\begin{table*}
\begin{center}
\caption{Parameters used for models displayed in Figures~\ref{fig1}--\ref{fig3}}
\label{tab1}
\scriptsize{
\vspace{-3mm}
\begin{tabular}{lccccc}
\hline
Property                &     Symbol       &  Unit    &       J1723   & J1227 \\  \hline
\multicolumn{5}{c}{Basic parameters} \\
Spin-down power  & $\dot E_{\rm SD}$       & $10^{34}\rm \ erg\ s^{-1}$   &    4.6  &   9.0   \\
Orbital period   & $P_B$                   &      day      &      0.615     &  0.288 \\
Reference Time   & $T_{\rm asc}$           &      MJD      &   55425.3205 & 57139.0716 \\
Orbital inclination   &  $i$                  &      degree      &      35     &  54.5 \\
Orbit radius   & $a_{\rm orb}$           &  $10^{11}$\,cm    &    2.6    &  1.5 \\
Companion mass   & $M_c$                   &  $M_\odot$    &      0.4    &  0.27 \\
Companion radius & $R_c$                   &  $R_\odot$    &      1       &  0.29 \\
Companion temperature     & $T_{\rm c}$          &     K         &     5300  &   5700  \\ \hline
\multicolumn{5}{c}{Pre-shock parameters} \\
Pair multiplicity in pre-shock wind & $\mathcal{M}$           &       $10^3$  &    2   &  2 \\
$e^-$ energy in pre-shock wind & $\gamma_w$              &       $10^4$  &  7    &    7    \\
Fraction of spin-down power in pre-shock wind & $\eta_w$             &              &   0.95     &    0.95    \\ \hline
\multicolumn{5}{c}{IBS parameters} \\
Momentum flux ratio      & $\beta$                 &               &    0.20 (0.36)$^{\rm a}$   & 0.16 (0.3)$^{\rm a}$ \\
Index for PL distribution of IBS $e^-$ & $p_1$                   &               &     1.3   &  1.3  \\
Minimum Lorentz factor of IBS $e^-$ & $\gamma_{s,\rm min}$    &               &       1    &  1  \\
Maximum Lorentz factor of IBS $e^-$ & $\gamma_{s,\rm max}$    &       $10^7$  &       1    &  1  \\
Fraction of spin-down power contained in IBS $e^-$  & $\eta_s$                &               &     0.89   &  0.92 \\
$B$ within IBS      & $B_s$                   &       G       &       2.66   &  1.9  \\
Bulk Lorentz factor of the IBS flow    & $\Gamma_{\rm D}$        &               &       1.2   &  1.5  \\
Length of IBS     & $l_{sh}$                   &$a_{\rm orb}$  &       4      &  4 \\ \hline
\multicolumn{5}{c}{Parameters for emission from the companion zone (shock-penetrating $e^-$)} \\
Companion $B$     & $B_c$                   &       kG      &       0.14   &  2.3 \\
Fraction of spin-down power in penetrating $e^-$   & $\eta_T$                &               &  0.95    &  0.95  \\
Penetration efficiency      & $\zeta$                 &               &       0.06    & 0.03 \\
Maximum Lorentz factor of penetrating $e^-$      & $\gamma_p$              &       $10^7$  &       7   &  7   \\
Size of the companion zone   & $l_{\rm comp}$          &$a_{\rm orb}$  &       4      &  4 \\ \hline
\end{tabular}}
\end{center}
\footnotesize{Note. For a detailed explanation of how these parameters are employed within the IBS model,
please refer to SAW24.\\
$^{\rm a}${Numbers in parentheses refer to the high-$\beta$ IBS, corresponding to the low X-ray flux LCs (red) in Figure~\ref{fig3} (see Sections~\ref{sec3_2} and \ref{sec3_3}).}
\vspace{-5 mm}
}
\end{table*}

\subsection{Modeling Long-term X-ray variability}\label{sec3_2}
\begin{figure*}[t]
\begin{tabular}{cc}
\hspace{7 mm}
\vspace{-2mm}
\includegraphics[width=2.7 in]{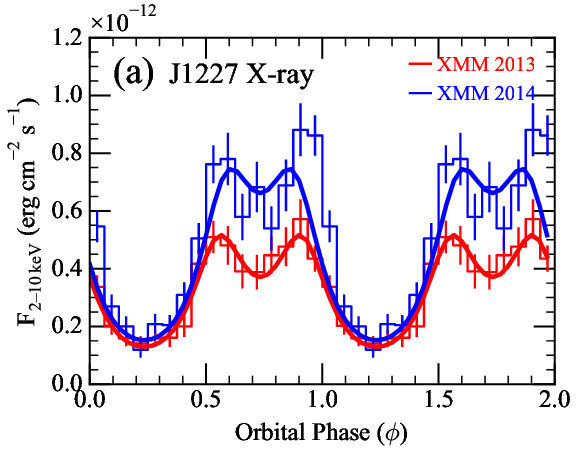} &
\hspace{7 mm}
\includegraphics[width=2.7 in]{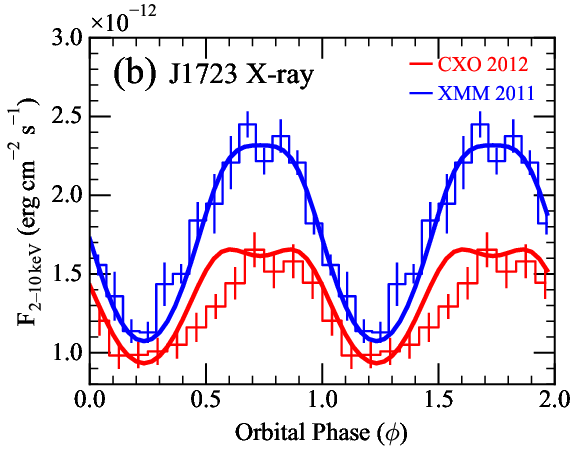} \\
\hspace{9 mm}
\includegraphics[width=2.7 in]{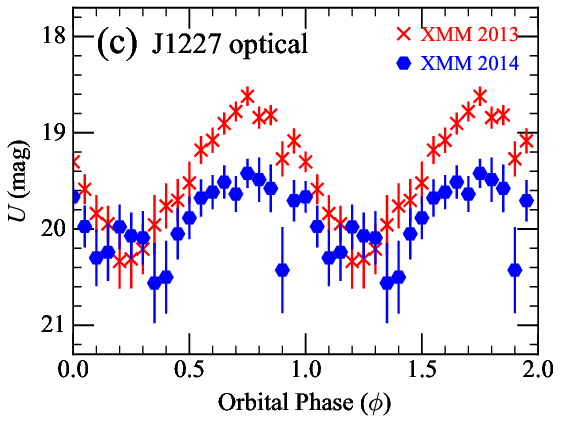} &
\hspace{14 mm}
\includegraphics[width=2.7 in]{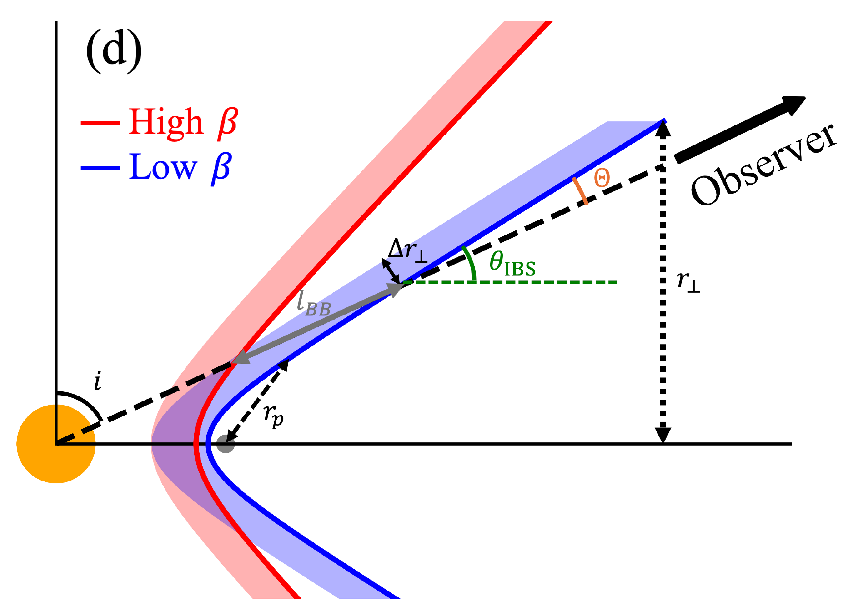} \\
\end{tabular}
\vspace{-5 mm}
\caption{\scriptsize{(a--b) 2--10\,keV LCs of J1227 (a) and J1723 (b)
measured by XMM-Newton and Chandra at different epochs \citep[][]{deMartino+15,Bogdanov+2014}.
Blue and red curves represent our IBS model.
(c) Optical (U band) LCs of J1227 corresponding to the X-ray LCs in panel (a).
(d) Schematic illustration of high-$\beta$ (red) and low-$\beta$ (blue) IBS configurations at $\phi=0.75$.
Only the low-$\beta$ IBS includes labels (e.g., $\theta_{\rm IBS}$ and $r_\perp$) for clarity.
Red and blue regions indicate stellar wind shocks for high- and low-$\beta$ cases, respectively.}
\vspace{-6 mm}
\label{fig3}}
\end{figure*}

Several RBs exhibit pronounced long-term variability in their X-ray emission.
Figure~\ref{fig3} presents X-ray LCs of the RBs J1227 (a) and J1723 (b), obtained with XMM-Newton
and Chandra at different epochs \citep[][]{deMartino+15,Bogdanov+2014}.
A significant variation in X-ray flux is evident between the two epochs for both sources.
In the case of J1227, the change in X-ray flux was accompanied by a corresponding
variation in the optical flux (Fig.~\ref{fig3} (c); Section~\ref{sec3_3}).

The observed X-ray variability implies changes in the properties
of the IBS formed by the interaction between the pulsar and companion winds \citep[e.g.,][]{ar17}.
Given the generally stable nature of pulsar winds,
variations in the companion's wind properties
are the likely cause of IBS changes.
This can affect the momentum flux ratio ($\beta$)
between the stellar and pulsar winds, defined as:
\begin{equation}
\label{beta}
\beta = \frac{\dot E_{\rm SD}}{\dot M_w v_w c},
\end{equation}
where $\dot M_w$ is the companion's mass-loss rate, $v_w$ is the wind velocity,
and $c$ is the speed of light. The parameter $\beta$ plays a crucial role in determining
the location of the IBS within the system (Fig.~\ref{fig3} (d)).

The parameter $\beta$ influences the X-ray emission through two primary mechanisms:
(1) a smaller $\beta$ positions the IBS closer to the pulsar, resulting in a stronger $B$
within the shock, as $B$ is primarily supplied by the pulsar and scales inversely
with distance from the pulsar to the emission zone ($r_p$; Fig.~\ref{fig3} (d)).
(2) The IBS opening angle $\theta_{\rm IBS}$ decreases with decreasing
$\beta$ values \citep[][]{Canto1996}:
\begin{equation}
\label{thetaIBS}
\theta_{\rm IBS} - \mathrm{tan}\theta_{\rm IBS} = \frac{\pi}{1-\beta}.
\end{equation}
The Doppler boosting of IBS emission depends on the angle
($\Theta$) between the IBS flow (approximated by $\theta_{\rm IBS}$) and the LoS.
Since the observer's inclination angle ($i$) remains relatively constant over long timescales,
variations in $\Theta$ ($\approx \theta_{\rm IBS} - \pi/2 + i$) are primarily driven by changes in $\beta$ through Equation~(\ref{thetaIBS}).
Our modeling indicates that the impact of $B$ on X-ray emission is more significant
than that of Doppler boosting due to the relatively low flow speeds ($c\sqrt{1-1/\Gamma_{\rm D}^2}$; Table~\ref{tab1})
within the IBSs of RBs and limited variation in $\theta_{\rm IBS}$.
Consequently, lower-$\beta$ IBSs generally produce brighter X-ray emission.

We applied our model to investigate the observed long-term X-ray variability
in J1227 and J1723. Using parameter values from Table~\ref{tab1},
we initially modeled the brighter (blue) LCs in Figure~\ref{fig3} (a) and (b).
To reproduce the lower-flux LCs (red in the figure), we adjusted the $\beta$
parameter (presented in parentheses in the table). As previously discussed, a strong stellar wind (low $\beta$) positions
the IBS closer to the pulsar, resulting in a higher $B$ within the shock and
enhanced X-ray emission. Conversely, a weaker stellar wind (high $\beta$) moves
the IBS away from the pulsar (Fig.~\ref{fig3} (d)), reducing X-ray emission.
To reproduce the measured LC variabilities, the model requires stellar wind variations
of a factor of $\lapp2$.

\subsection{Simultaneous Optical and X-ray Variability in J1227}\label{sec3_3}
Another intriguing aspect revealed by XMM-Newton's simultaneous X-ray and optical
observations of J1227 is the observed anti-correlation between the long-term fluxes
in these two wavebands (Fig.~\ref{fig3} (a) and (c)).
\citet{deMartino+15} proposed that long-term variations in optical emission
could be linked to changes in shock size. A larger shock might produce
stronger X-ray modulation while simultaneously causing extinction of some optical light from
the companion. However, our IBS model presents challenges for this scenario,
as a wider IBS would typically result in weaker X-ray emission.

We propose an alternative explanation for the anti-correlated X-ray and optical variability
observed in J1227. Rather than attributing this behavior to the shock region blocking
the optical emission, we suggest that the depth ($l_{BB}$) of the stellar-wind shock
traversed by the companion's blackbody photons is crucial (Fig.~\ref{fig3} (d)).
The parameter $l_{BB}$ reaches its maximum at $\phi=0.75$ (optical maximum) and is 0 at $\phi=0.25$.
For a given shock thickness ($\Delta r_\perp$), $l_{BB}$ scales with
$\Theta$ as $l_{BB} \mathrm{sin}\Theta \approx \Delta r_\perp$.
The number of hydrogen atoms injected during a particle residence time $\tau_r $ ($\approx l_{sh}/v_w$ with $l_{sh}$ being the IBS length) within the shock
is $N_{\rm H}=\dot M_w \tau_r f_\Omega/m_{\rm H}$, where $f_\Omega = (1 - \mathrm{cos}\theta_{\rm IBS})/2$ takes into account the solid angle subtended by the shock from the companion,
and $m_{\rm H}$ is the hydrogen mass.
The volume of the stellar-wind shock is approximated as $V_{S}\approx \pi r_{\perp} l_{sh}\Delta r_{\perp}$
with $r_{\perp} = l_{sh}\mathrm{sin}\theta_{\rm IBS}$ (assuming a cone).
The neutral hydrogen column density along $l_{BB}$ can then be estimated as
\begin{equation}
\label{nH}
n_{\rm H} = \frac{N_{\rm H} l_{BB}}{V_{S}} \approx \frac{\dot M_w f_\Omega}{\pi m_{\rm H} v_w l_{sh}\mathrm{sin}\theta_{\rm IBS} \mathrm{sin}\Theta}.
\end{equation}

Equation~(\ref{nH}) demonstrates that variations in the companion's wind (i.e., $\dot M_w v_w$) can substantially affect the extinction \citep[$A_V\propto n_{\rm H}$;][]{Diplas1994} of the its emission. This is because the wind influences $\beta$ (Eq.~(\ref{beta})), $\theta_{\rm IBS}$ (Eq.~(\ref{thetaIBS})) and $\Theta$.
For small values of $\Theta$ (e.g., Fig.~\ref{fig3} (d)), the $\mathrm{sin}\Theta$ term dominates. Consequently, even a slight change in $\Theta$ can lead to a substantial alteration in extinction. Then, the large change in $\beta$ (and $\Theta$) inferred from the X-ray variability in J1227 between 2013 and 2014 (Table~\ref{tab1} and Section~\ref{sec3_2}) offers a qualitative explanation for the observed difference in the U-band magnitudes at $\phi=0.75$ (Fig.~\ref{fig3} (d)) between the two epochs, although changes in individual $\dot M_w$ and $v_w$ may complicate the estimation of $A_V$.
The consistent minimum optical flux (at $\phi=0.25$) in both years
further supports this scenario, as the shock does not obstruct the companion at that phase.
Moreover, this model naturally explains the observed anti-correlation between the long-term
optical and X-ray flux variations.

\section{Discussion and Summary}\label{sec4}
We refined the IBS model of SAW24 by incorporating the dynamic evolution of electrons within the shock due to radiative cooling. Consequently, the electron spectrum exhibits a lower cutoff energy than that predicted by the SAW24 model, yet this cutoff remains beyond the reach of current high-energy observatories. As a result, this refinement had a negligible impact on
the previously reported results for three RBs. Similar to the SAW24 model, our model successfully explains the high-energy emission
from the RBs if pulsar wind electrons possess energies of $\sim$40\,TeV,
penetrate the IBS, and interact with the companion's magnetosphere.  While the required voltage drop
\citep[$\Delta V\approx 6.6\times 10^{12} B_{12}P^{-2}$\,volts;][]{Ruderman1975} is achievable
by the millisecond pulsars in those RBs, further theoretical investigation into the dynamics
and propagation of such high-energy particles to the companion region is necessary.

Our model results for J1723 are very similar to those of \citet{Merwe+2020},
who attributed its X-ray (and GeV) emission to synchrotron radiation from the IBS.
However, a significant discrepancy arises in the TeV band,
where our model predicts an order of magnitude lower flux than \citet{Merwe+2020}.
This difference stems from our assumption of a larger IBS length,
resulting in a greater average distance between TeV-emitting electrons
and the companion, and consequently a lower seed photon density within
the emission region. Future TeV observations can differentiate between
these models, providing crucial insights into the IBS length.

We successfully modeled the long-term X-ray variability observed in J1227 and J1723
by attributing it to variations in the $\beta$ parameter driven by changes in the stellar wind.
These changes also induce variation of $l_{BB}$ (and $A_V$), causing variability in the observed optical flux. While this simplified model captures the main features of the observed long-term X-ray
and optical variability, a more comprehensive treatment incorporating the full IBS geometry
is required to accurately fit the optical light curves beyond $\phi=0.75$.
Additionally, irradiation feedback \citep[][]{Wadiasingh2018} and physical conditions within the stellar-wind shock can influence the results. The phenomenological IBS model presented here provides a foundation for understanding optical variability, but further investigation using physics-based models is warranted. Such a model can be further tested through both optical photometry and spectroscopy.

\section*{Acknowledgments}
JP acknowledges support from Basic Science Research Program through the National
Research Foundation of Korea (NRF) funded by the \fundingAgency{Ministry of Education} (\fundingNumber{RS-2023-00274559}).
This research was supported by the National Research Foundation of Korea (NRF)
grant funded by the \fundingAgency{Korean Government (MSIT)} (\fundingNumber{NRF-2023R1A2C1002718}).





\bibliography{redbacks}%


\end{document}